\documentclass[a4paper,10pt,twocolumn,english]{article}

\usepackage{amssymb}
\usepackage{amsmath}

\usepackage{inputenc} 
\usepackage{graphicx} 
\usepackage{multirow} 

\title{Visualization and analysis of jet oscillation under transverse acoustic perturbation \\
}
\author{
\normalsize{ Patricio de la Cuadra}\\%
	\footnotesize{Centro de Investigaci\'on en Tecnolog\'ias de Audio (CITA)
Universidad Cat\'olica de Chile, 
	e-mail: \textbf{pcuadra@uc.cl}}\\%
\normalsize{\bf Christophe Vergez} \textbf{(corresponding author)}\\%
	\footnotesize{LMA, CNRS, 31 Chemin Joseph Aiguier, 13402 Cedex 20 Marseille, France, 
	e-mail: \textbf{vergez@lma.cnrs-mrs.fr}}\\%
\normalsize{Benoit Fabre}\\%
	\footnotesize{LAM-IJLRDA, Paris 6 University, 11 rue Lourmel, 75015, Paris, France, 
	e-mail: \textbf{fabreb@ccr.jussieu.fr}}\\%
}

\begin{document}
\maketitle
\thispagestyle{empty}
\begin{abstract}

Schlieren flow visualizations of transverse oscillations of jets submitted to an
acoustic perturbation are analyzed in this paper. The aim is to estimate the shape and the position of the median line of the jet.
Two methods for
image processing are proposed, based on complementary approaches~:
inter image and intra image analysis.  Synthesized  images of an oscillating jet are used to validate each method and compare their performances in the case of noisy pictures. Illustrations are then shown on real laminar and turbulent jets. The results obtained using both methods  are
very close, showing their reliability. Applications investigated in this paper are focused on the estimation of the convection velocity of perturbations along the jet, and the influence of the Reynolds number and of the channel geometry upstream the jet formation.
\end{abstract}


\section{Introduction}

In flute-like instruments, acoustic oscillation is generated by the
coupling of the unstable modes of an air jet and the acoustic modes of
a pipe resonator. For soft blowing conditions, the jet is laminar and
the instability is usually described, following Rayleigh's approach (\cite{rayleigh45_2}),
under assumptions of infinite inviscid 2-D flow. Certainly this
approach needs to be combined with ad-hoc assumptions to cope with the
specific geometry of the problem (semi-infinite jet, acoustic
transverse perturbation). Furthermore, because it is linear, this perturbation
approach may be limited to small amplitudes of jet deflection. For many
instruments like transverse flutes and popular music flutes, the sound
power required calls for higher total jet flux. Reynolds numbers,
based on jet thickness and velocity, up to 10000 have been estimated
in transverse flutes. Under such conditions, the jet becomes rapidly
turbulent downstream from the flue exit. The development of
instability induced by the acoustic perturbation on such jets has been
studied by Bechert (\cite{bechert76}, \cite{bechert88}) and Fletcher (\cite{thwaites80}, \cite{thwaites82}). Both propose semi-empirical models to
describe the jet oscillations.

The intimate detail of the coupling between the acoustic perturbation and the jet instability remain an
unsolved problem. It appears that for both laminar and turbulent jets,
the lack of theoretical models make experimental data useful to fit,
calibrate and evaluate semi-empirical models (\cite{fabre2000a}).

Different researchers have studied experimentally jet oscillations
under acoustical perturbations using hot-wire anemometry measurements (\cite{nolle98}, \cite{yoshikawa98} for example). On the other hand, flow visualization has
been  widely used as an inspiration for a physical analysis and
description of the fluid mechanisms and aeroacoustics at work in
flutes  (\cite{verge94}).

In the present paper, we present tools for image analysis of flow
visualizations to extract quantitative data on jet instability under
acoustic perturbation. The development of instability on a jet is
experimentally investigated using a simplified device (see section \ref{s:device}): the pipe
resonator is removed from the flute and replaced by an acoustic field
generated by loudspeakers. Using the Schlieren technique, flow visualizations have been carried
out for a variety of Reynolds and Strouhal numbers.

Two different image analysis techniques have been developed (see
sections \ref{s:imag_proc} and \ref{s:valid}) and they
are discussed in detail in this paper. 
The extracted data is analyzed in
the framework of a harmonic perturbation that is convected downstream
and grows in the shear layers of the jet (see section \ref{s:results}).

Finally, the influence of the Reynolds number and the channel geometry on the jet behavior are presented as an application in section \ref{s:disc}.

\section{Experiment description \label{s:device}}

\begin{figure}[h]
\center
\includegraphics[width=.6\columnwidth]{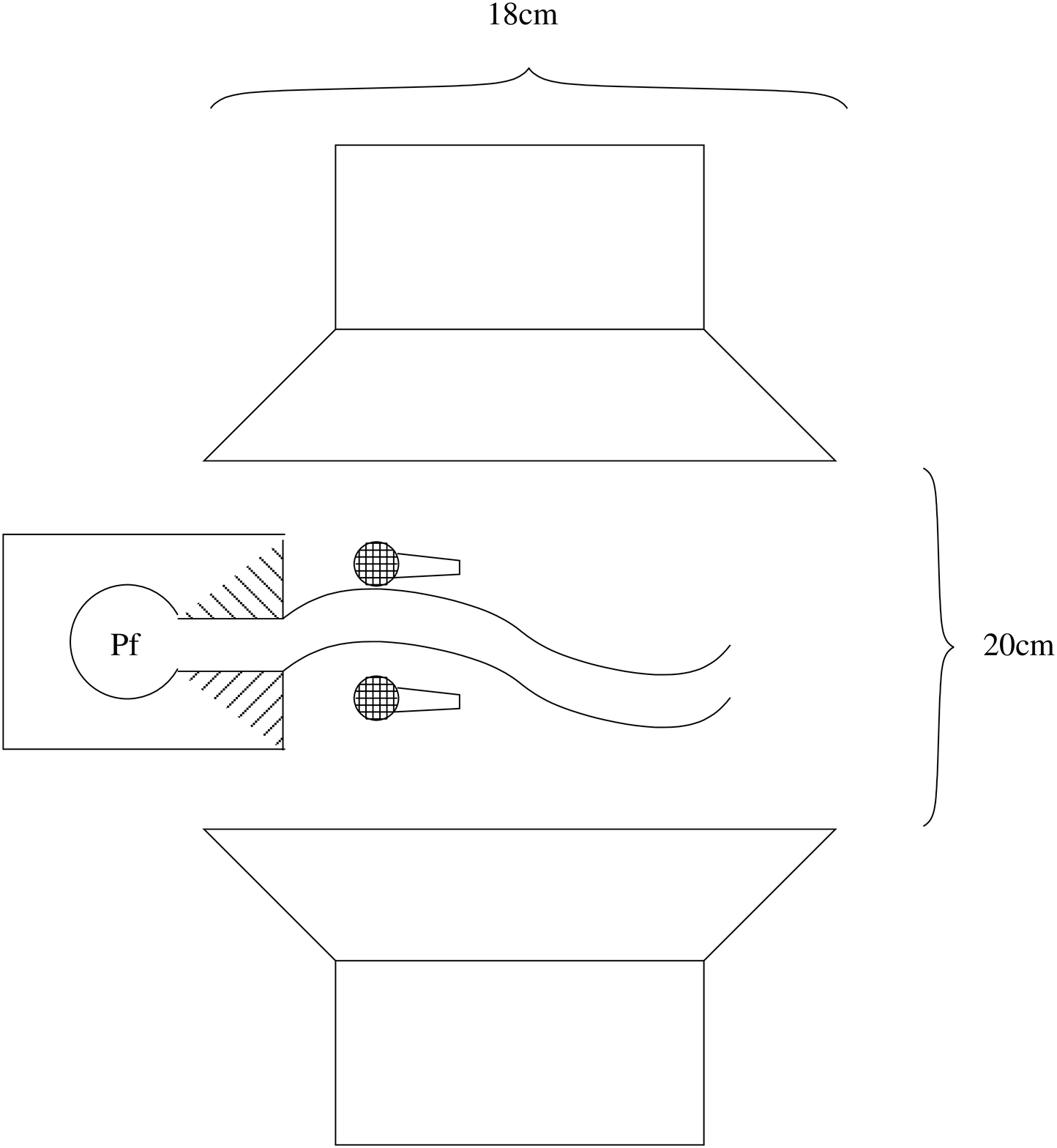}
\caption{Experimental setup. Jet exit, speakers and two microphones (acoustic velocity estimation).}
\label{fig:manip_speaker_sensor}
\end{figure}

\paragraph{Experimental setup:} a jet, created by blowing through a slit, is acoustically forced by
two out-of-phase loudspeakers (see figure \ref{fig:manip_speaker_sensor}). 
The speed of the jet is controled by varying the pressure in the cavity $p_f$ just before the jet formation. The jet central speed at the channel exit  $U_b$, can be estimated using Bernoulli's equation:
\begin{equation}
U_b = \sqrt{\frac{2 p_f}{\rho_0}},
\label{eq:Bernoulli}
\end{equation} 
where $\rho_0$ is the fluid density.

\paragraph{Jet structure:} two Reynolds numbers $R_e = 500$ and
$R_e = 3000$  were chosen ($R_e \triangleq U_bh/\nu$, with $h=10^{-3}$m the channel height
and $\nu$ the kinematic viscosity). They correspond to two different
operating modes of the jet: $Re = 500$ gives a velocity around 7.5 m/s
which corresponds to a normal blowing condition for a recorder and
assures a laminar behavior, while $Re = 3000$ corresponds to a velocity around
45 m/s producing a turbulent jet.

 The acoustic excitation frequency $f$
is ranging from 70 to 1400 Hz (when $Re = 500$) and from 1200 to
2900Hz (when $Re = 3000$) to cover a Strouhal ($S_t$) range
from approximately 0.05 to 1.15 ($S_t \triangleq fh/U_b $). A fixed amplitude of the
acoustic velocity  is chosen for each of the two experiments described in the present paper
and a typical value is $0.5\%$ of the jet velocity\footnote{This is considerably lower that the $10\%$ observed in real flutes but allows us to cover an interesting range of frequencies without over exciting the speakers.}. The acoustic velocity is measured using a Microflown velocity sensor (low amplitudes) and a microphone doublet ( high amplitudes) at a position near the unperturbed jet path, but out of the flow visualization window. Preliminary experiment had been carried out to check the acoustic velocity homogeneity in the jet and sensor area. 

\begin{figure}[h]
\center
\includegraphics[width=.9\columnwidth]{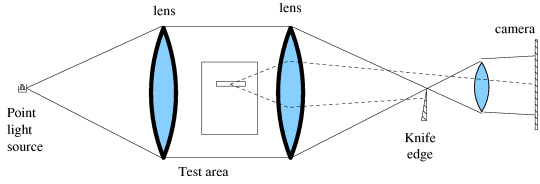}
\caption{Schlieren basic implementation: a subset S of the test area
  with different medium (here the $\textrm{CO}_2$  jet) deviates light rays. A knife edge, placed in the
  focus of the second lens blocks part of the rays having crossed S,
  thus producing intensity variations on the screen (see figure \ref{fig:RealJet}).}
\label{fig:lens_array} 
\end{figure}

\begin{figure}[h]
\center
\includegraphics[width=.9\columnwidth]{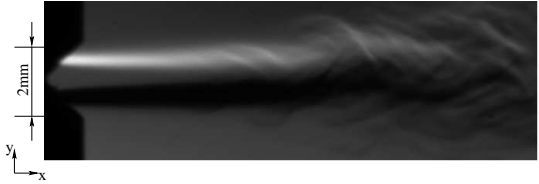}
\caption{Schlieren jet image (see figure \ref{fig:lens_array} for the
  optical principle).}
\label{fig:RealJet} 
\end{figure}

\paragraph{Schlieren jet images:}
Schlieren technique
(\cite{merzkirch87}, section 3.2.2 p134) is used to observe the behavior of the oscillating jet: using an optical scheme, light phase shift crossing an inhomogeneous media is converted into light intensity (see figure \ref{fig:lens_array}).  We use
$\textrm{CO}_2$ jet traversing the air, whose mass
  density assures enough difference to allow Schlieren
  visualization (see figure \ref{fig:RealJet}) while producing a behavior close to that found in
  real flute-like instruments. Sequential images of the jet are taken with a digital camera
synchronized with a stroboscopic lighte. 
 Its frequency is set such that multiples of its frequency are slightly out of phase
  with the excitation frequency providing an aliased representation of
  the oscillation. Frequency of the camera was set to
  $14$fps, exposure time to $1\mu s$. Approximately $100$ images are taken covering
two cycles of jet oscillation. Images are captured in raw, black and
white, bmp files with size 1280 x 448, and 8 intensity bits.

\section{Image processing \label{s:imag_proc}}

The goal of the image processing algorithm is to detect transverse jet
displacement 
for every image in the oscillation period.
Two algorithms are proposed in this section. Their performances will be checked and compared in section \ref{s:valid} and \ref{s:results} respectively.

\subsection{Cross-correlation method}

A first algorithm based on cross-correlation between successive images
is proposed.

\begin{figure}[h]
\center
\includegraphics[width=.49\columnwidth]{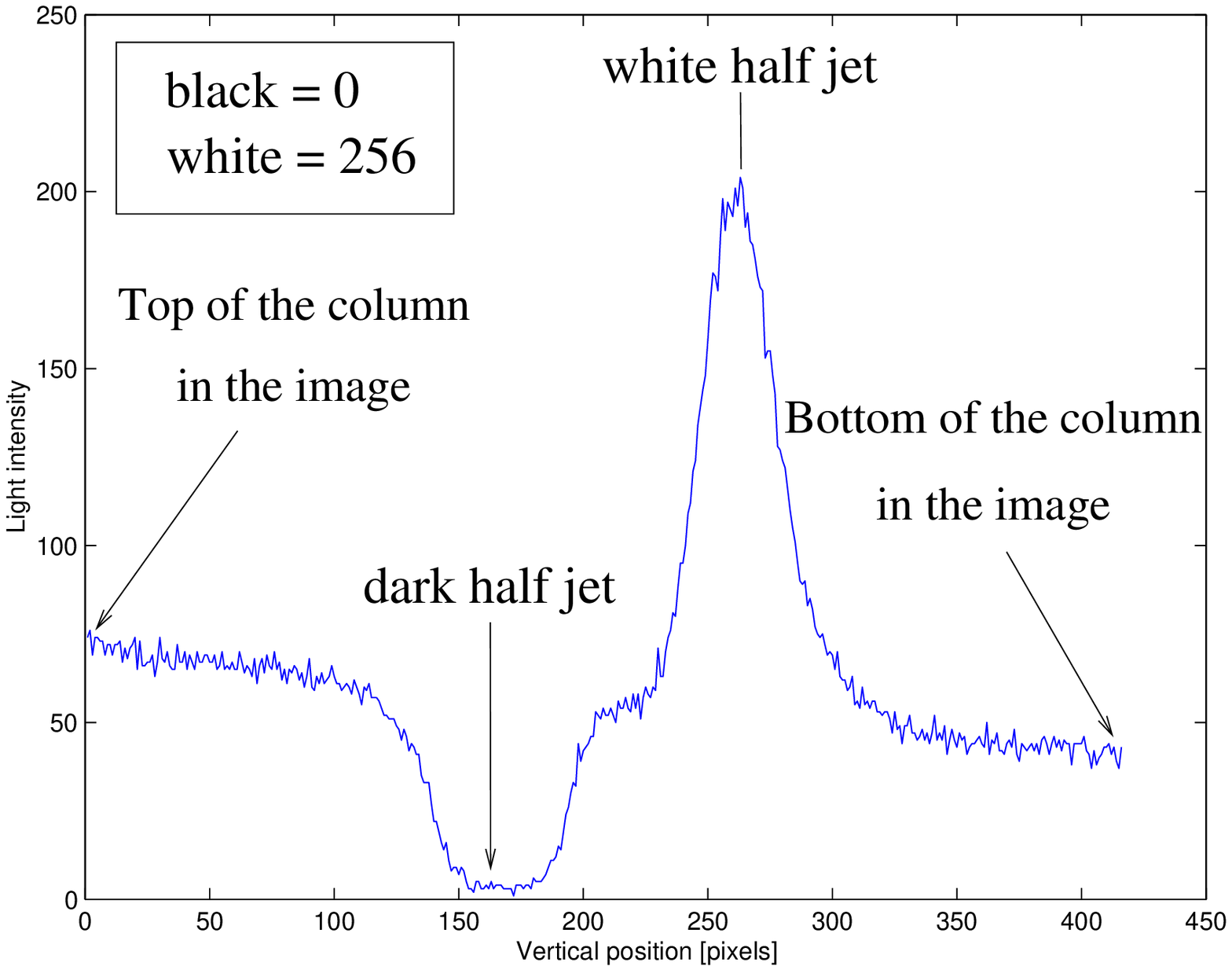}
\includegraphics[width=.48\columnwidth]{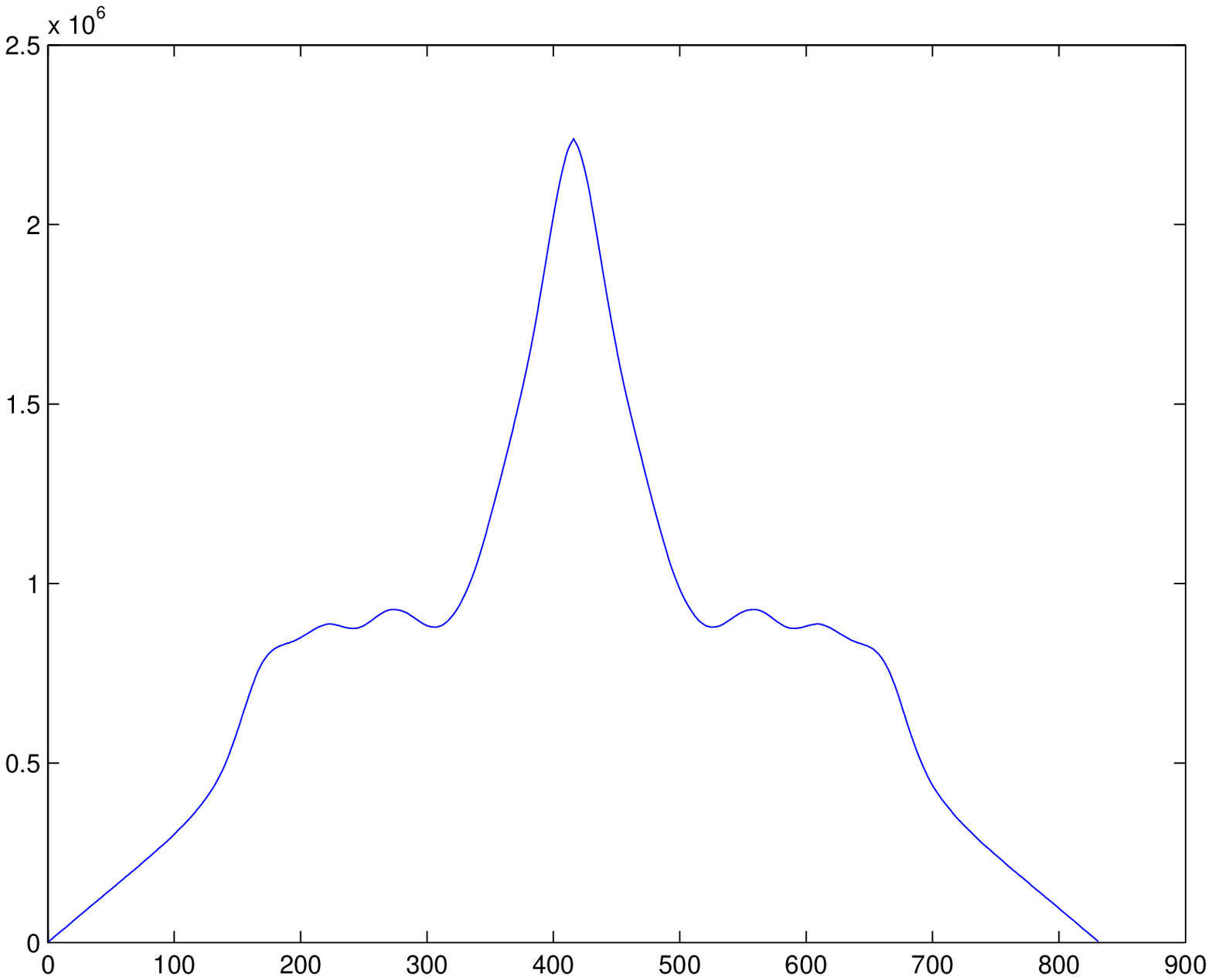}
\caption{Intensity level of a single column from a Schlieren image
  (left) and cross-correlation of two intensity graphs corresponding
  to the same
  column in two successive images of the sequence (right).}
\label{fig:intensity} \label{fig:crosscorrelation} 
\end{figure}

Let us consider the intensity graph of one column of the image as shown in
figure \ref{fig:intensity} (left picture). For that particular column,
the shape of the graph does not change much as time goes on: 
the shape is only shifted, according to the transverse position $\eta$ in the $y$ direction due to the movement of the jet. However, the
particular shape varies with the column considered since it is determined by the
mass distribution of the jet as shown in figure \ref{fig:RealJet}.

Cross-correlation is calculated between two intensity graphs
corresponding to the same column of two
successive images of the sequence (see right part
of figure \ref{fig:crosscorrelation}).  The position of the peak allows to measure the displacement
of the jet between these two successive images. This is repeated for each column of the image.
Since changes on the jet position may be smaller than one pixel,
a parabola is fitted to the three highest points to estimate the
position of the peak. When considering all the columns of all the
images in the sequence, the jet position can be reconstructed as shown
in figure \ref{fig:Position_mesh_crosscor} ($R_e = 500$ and $S_t = 0.2$).
\begin{figure}[h]
\center
\includegraphics[width=0.9\columnwidth]{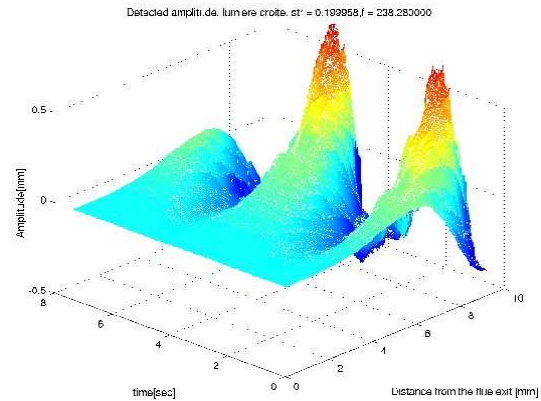}
\caption{Jet deviation, as estimated after  image
  analysis through cross-correlation approach, in the case $R_e = 500$ and $S_t = 0.2$}
\label{fig:Position_mesh_crosscor} 
\label{fig:Position_mesh_morpho}
\end{figure}

 Some defects of the
experimental setup are clearly visible in left part of figure
\ref{fig:intensity}: the two peaks corresponding to the brighter and
darker part of the image  should  ideally be of the same height,
but in practice a symetrical contrast is very difficult to obtain. Moreover, unhomogenous background intensity
(inside each image) is represented as a non horizontal line on the borders of the left image in figure \ref{fig:intensity}. However, these features of the
images are useful for the cross-correlation since they help to produce the well
shaped peak in the curve observed in the right part of figure \ref{fig:crosscorrelation}.

\subsection{Morphological method}

The idea of this method is to apply morphologic functions on binarized
images, in order to identify large spatial scales, corresponding to the jet main structure, while ignoring the smallest scales (vortices, background noise ...).  This is done
through the five following steps (also illustrated in figure \ref{morpho_illust}):

\paragraph{Contrast enhancement and homogenization between all the images
  of a sequence (figure \ref{morpho_illust}b):} the contrast is first enhanced by histogram
  equalization; the histogram of each image is streched to lie on the
  full 8-bit range, and to match a flat histogram. This also plays the role of contrast homogenisation between different images of a given sequence. 

 \paragraph{Conversion from greyscale to black and white images (figure \ref{morpho_illust}c)}: a
  statistical method based on the Otsu principle estimates the threshold which best separates the histogram within
  two classes (\cite{otsu79}). The binarization threshold is therefore
  different for each image. As a global method, the efficiency of the
  Otsu method is known to vary according to the type of image. A classical improvement is adopted, which consists in multiplying the
  Otsu threshold by an empirical constant in order to better enhance
  the jet structure for a given sequence of image. Typical values for this constant lie beween 0.9 and 1.1. In the resulting  BW images, the jet is represented by a lower and a upper half jet, black and white respectively (figure \ref{morpho_illust}c). In the following, both half jets are treated separately~: the same operations are applied to the BW image and to its 1-complemented image. For clarity, only operations on the image containing the lower half jet are shown in figures \ref{morpho_illust}d,  \ref{morpho_illust}e and  \ref{morpho_illust}f. 

\paragraph{Extraction of largest spatial scales  in the image ( figures \ref{morpho_illust}d and  \ref{morpho_illust}e):} for each half-jet, the black and white picture includes several black structures, among which the half jet. This image is usually noisy, with black dots
  or small-sized sets of black pixels on the background. This can
  result either from a non-optimal threshold, or from the Schlieren
  experiment itself, which does not produce perfectly uniform
  backgrounds (mainly because the optical lens are never dispersion free). Therefore, a morphological closure (\cite{niblack86}) is performed to
  clean all small structures  (figure \ref{morpho_illust}d). Then a
  morphological opening is applied in order to highlight large scales
  in the image  (figure \ref{morpho_illust}e). The structuring element is a disk, the size of which (optimized manually), is the same for all the images of a given sequence.

\paragraph{Identification of the half-jet among the remaining large scales (figure \ref{morpho_illust}f):}
The image is decomposed into different regions, defined as groups of contiguous pixels. A unique region, corresponding to the estimated half-jet, must be retained. However, since the jet is oscillating, sometimes with large amplitudes, its position may vary significantly between different images of the same sequence. Therefore the discriminating criterion retained, involves both the area and the localization of each region and uses the (known) main flow direction from left to right).

\paragraph{Jet edges and median line estimation}: jet edges
  detection in figure \ref{morpho_illust}f is trivial and is performed with Sobel algorithm  which
  uses gradient information (\cite{duda73}). Edge detection is interesting in order to
  analyse in further studies the spreading of the jet as a function of the distance from
  the exit. Moreover, it is an intermediate step to estimate the
  median line, which we are mainly interested in. Finally, an image containing edges of both half jets is scanned column after column and the median line is constructed point after point, including empirical criteria to select, among several possible paths, the one to follow. This guarantees that the resulting median line is single-valued, but possibly discontinuous in the case of vortices.

\paragraph{Structure length:} The largest the distance from the exit, the most complex the structure of the jet, and the least reliable the estimation of the median line. An index of confidence is then used: the estimation of the median line stops at the distance (i.e. the column) corresponding to the length of the smallest half-jet. Indeed beyond that limit, median line estimation would rely on information given by a single half-jet (as can be seen in figure  \ref{im_med_line}). This distance depends on the experimental conditions (Reynolds and Strouhal numbers) but obviously varies from an image to another within a sequence, depending both on the phase of the excitation (whether the jet is horizontal or deflected), and on the image itself (the whole process of jet identification being more or less efficient from an image to another).

 The result of this method on one image is presented in figure
 \ref{im_med_line}. Bringing together the median line found for each
 image allows to construct 3D plots similar to those obtained
 through cross-correlation image analysis (figure
 \ref{fig:Position_mesh_morpho}).

\begin{figure}
\addtocounter{figure}{1}
\begin{tabular}{cc}
\begin{tabular}{c}
{\scriptsize (\thefigure a) Original image} \\ 
\includegraphics[width=.42\columnwidth]{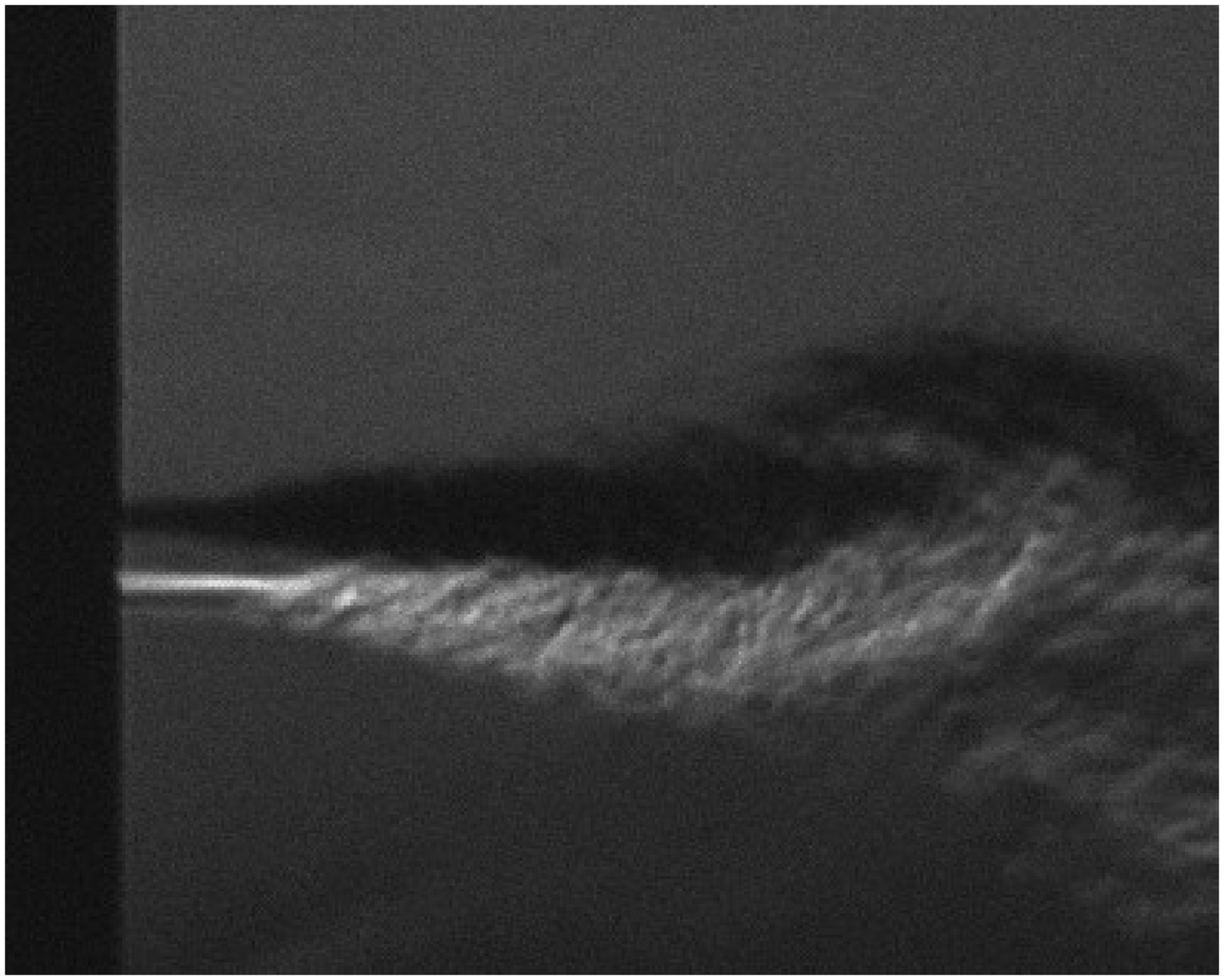}
\end{tabular}
&
\begin{tabular}{c}
{\scriptsize  (\thefigure b) Contrast enhanced} \\ 
\includegraphics[width=.42\columnwidth]{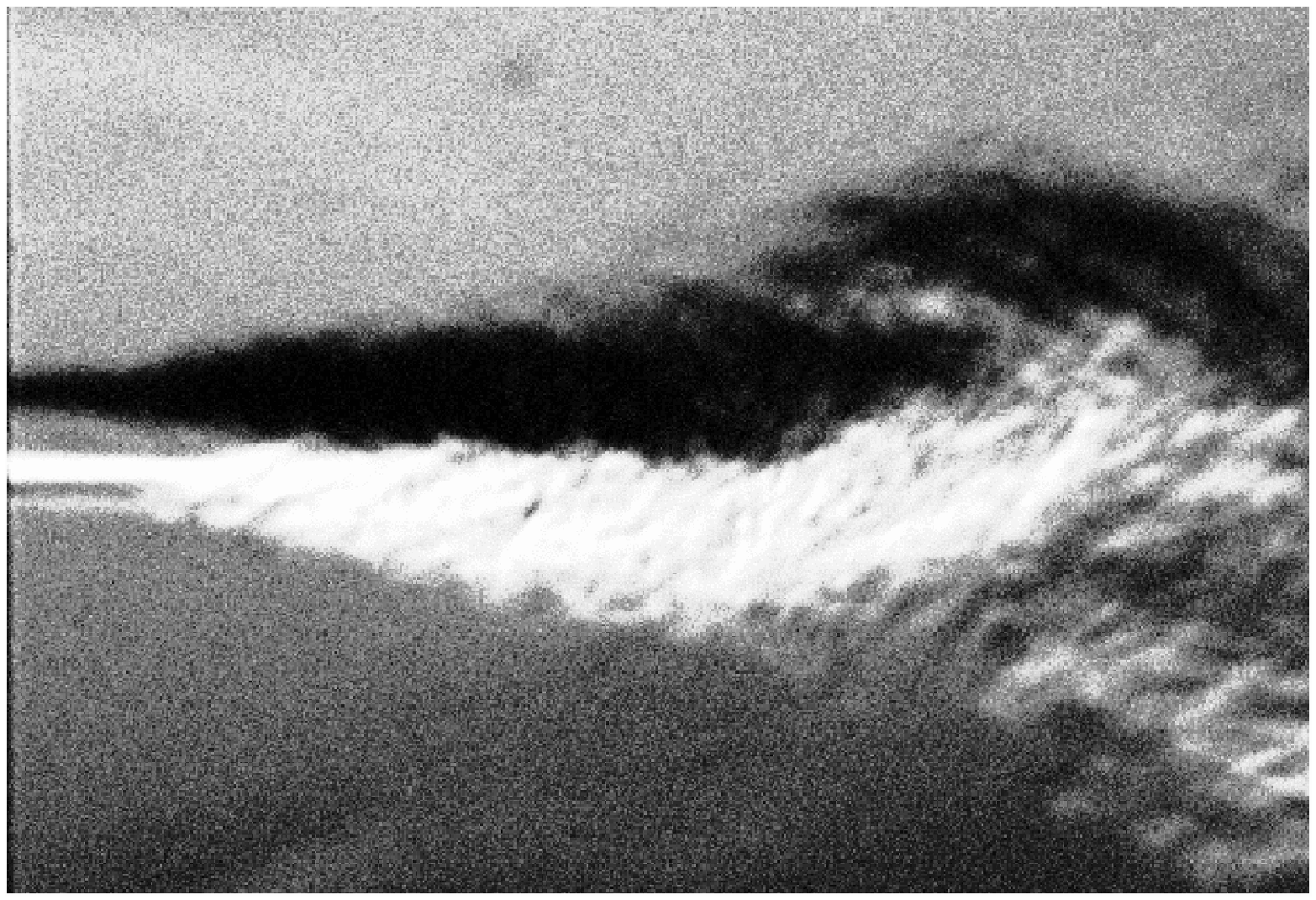} 
\end{tabular}
\\
\begin{tabular}{c}
{\scriptsize   (\thefigure c) Automatic binarization} \\
\includegraphics[width=.42\columnwidth]{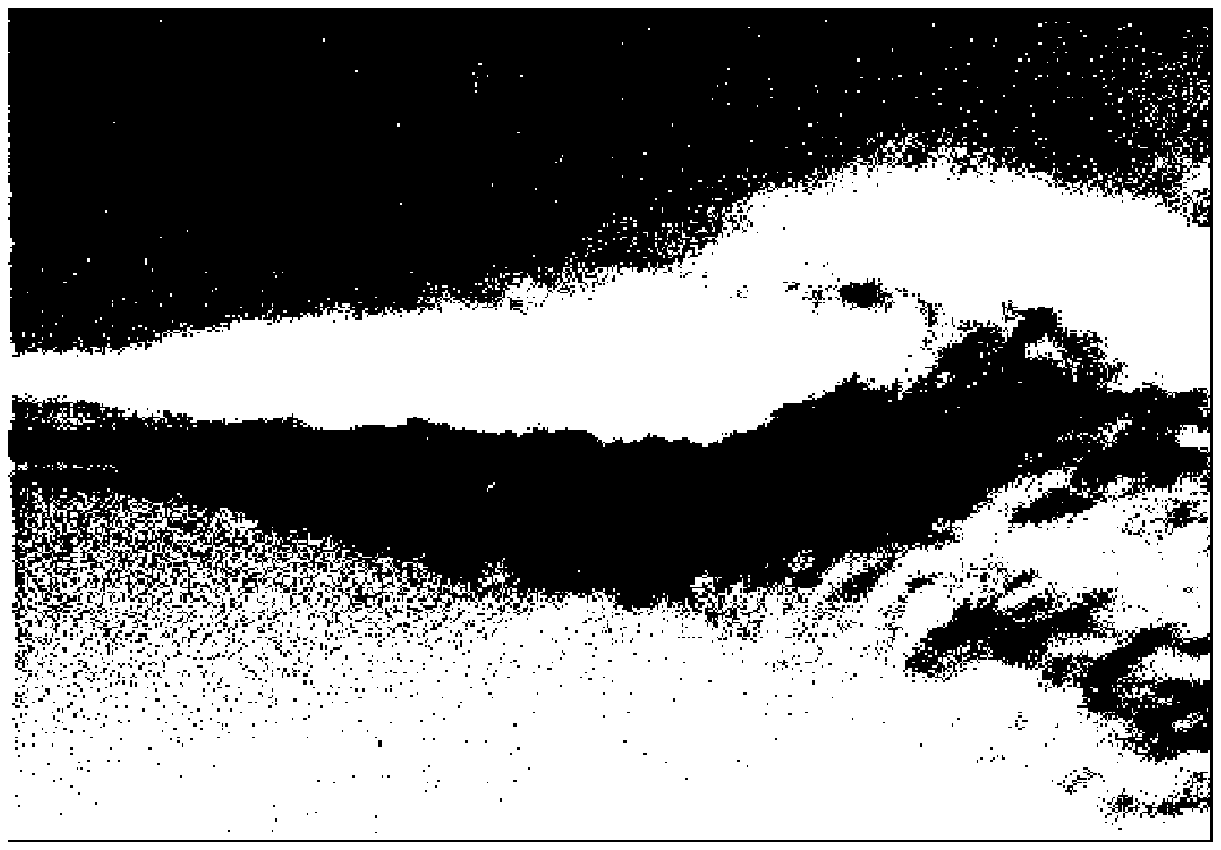}
\end{tabular}
&
\begin{tabular}{c}
{ \scriptsize (\thefigure d)  Morpho. closure} \\
\includegraphics[width=.42\columnwidth]{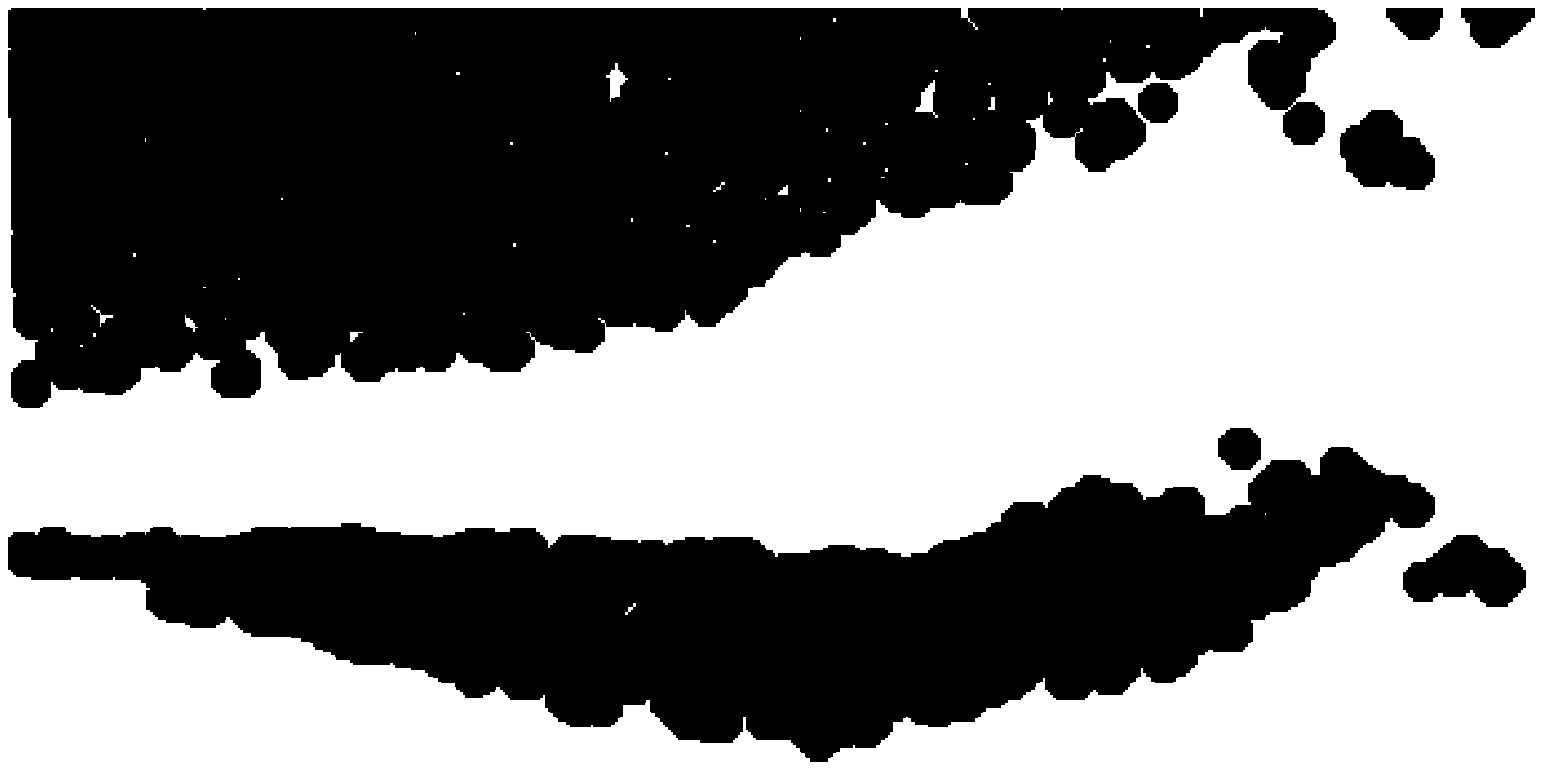} 
\end{tabular}
\\
\begin{tabular}{c}
{ \scriptsize  (\thefigure e) Morpho. opening} \\
\includegraphics[width=.42\columnwidth]{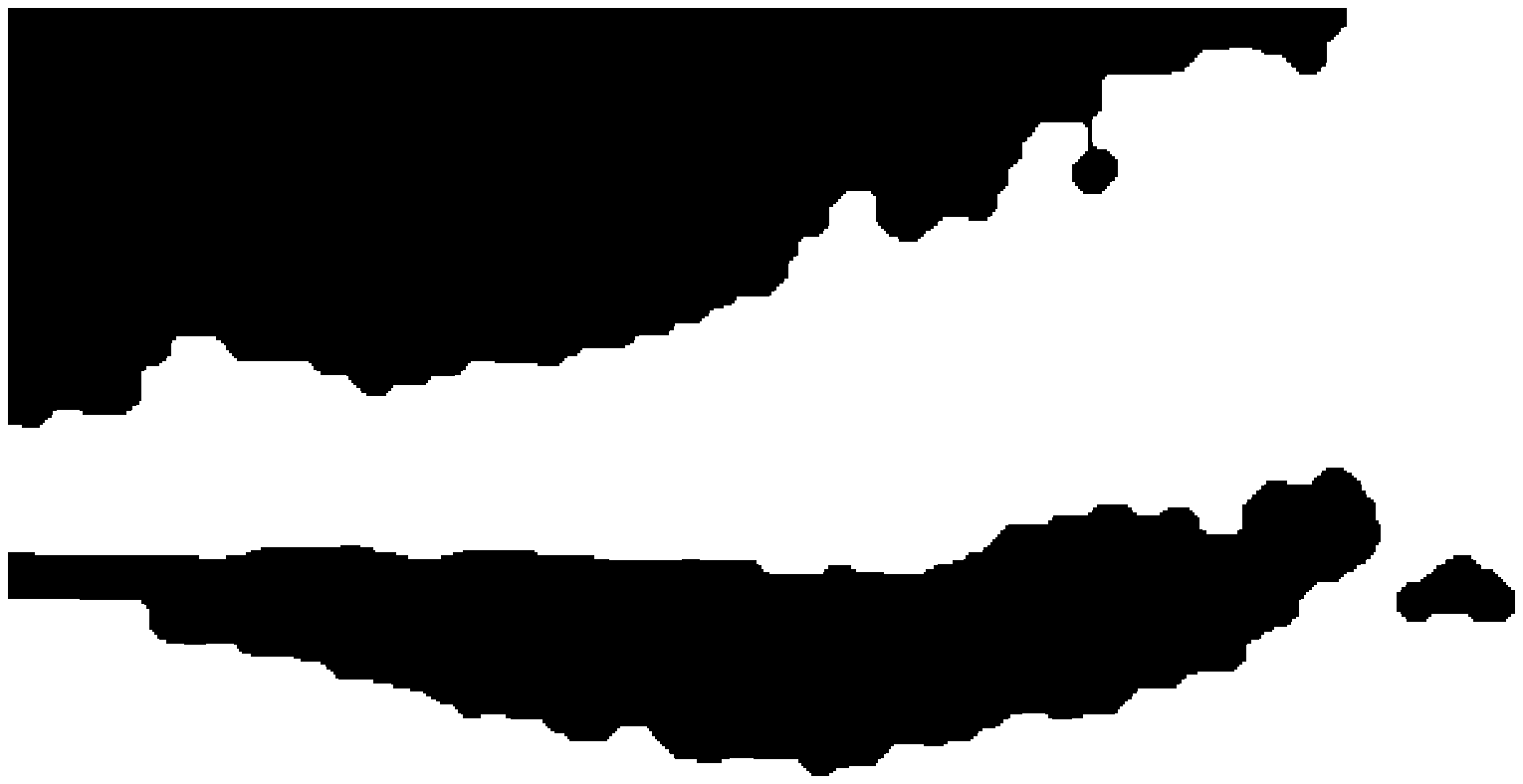}
\end{tabular}
&
\begin{tabular}{c}
{\scriptsize   (\thefigure f) Jet identification} \\
\includegraphics[width=.42\columnwidth]{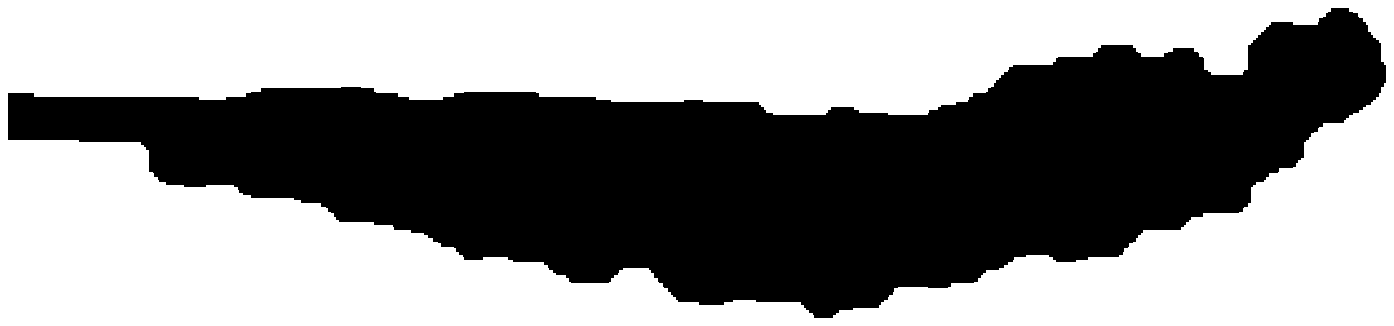} 
\end{tabular}
\end{tabular}
\addtocounter{figure}{-1}
\caption{Morphological method: illustration
  of the different steps (from left to right and top to bottom). \label{morpho_illust}}
\end{figure}

\begin{figure}[h]
\center
\includegraphics[width=.9\columnwidth]{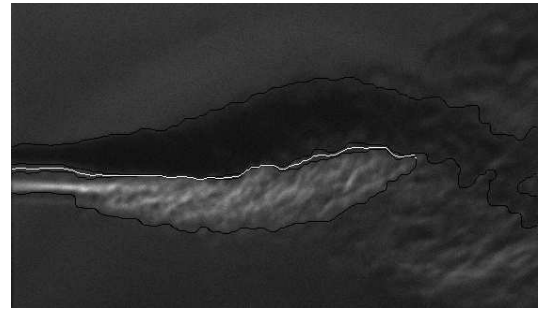}
\caption{Original Schlieren image superimposed with edge detected
  (black line) and the median line deduced (white line). \label{im_med_line}}
\end{figure}

\section{Evaluation of both methods \label{s:valid}}

Both methods are evaluated in this section: they are first validated through the use of computer made images  (section \ref{ss:valid})  and their robustness against background noise is compared in section \ref{ss:noise}.

\subsection{Validation with digitally synthesized jet movies \label{ss:valid}}
To validate both methods, the image analysis procedures described in
section \ref{s:imag_proc} are applied to digitally synthesized images of a
jet (see figure \ref{f:synth_jet_noise_influence}). 

The images have been synthesized to imitate Schlieren images of real
laminar jets like the ones captured in our experiments: same lines and columns numbers, 8-bit
intensity images, upper and lower shear layers of the jet
respectively  black and white, grey homogenous background.

The synthesized jet has an exponentially amplified sinusoidal motion
with controlable parameters.
It is checked for different parameters of the jet (jet width,
amplification coefficent, frequency of oscillation) that both methods
are sucessful (see for example a graphical confirmation in the first row of figure
\ref{f:synth_jet_noise_influence} where the median line is superimposed with the synthesized image).

\subsection{Sensitivity to background noise  \label{ss:noise}}
\begin{figure*}
\addtocounter{figure}{1}
\begin{tabular}{cc}
\begin{tabular}{c}
{\scriptsize   (\thefigure e) -3dB noise added (morphological analysis)} \\
\includegraphics[width=.95\columnwidth]{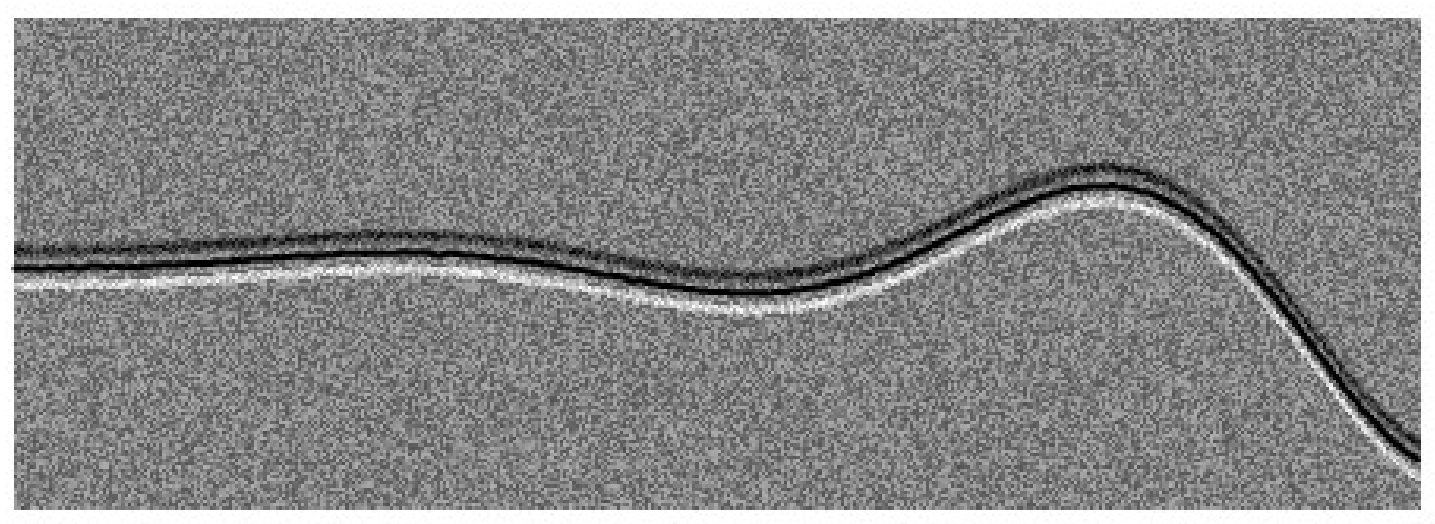}
\end{tabular}
&\hspace*{-1.cm}
\begin{tabular}{c}
{ \scriptsize (\thefigure f)  -3dB noise added (crosscorelation analysis)} \\
\includegraphics[width=.95\columnwidth]{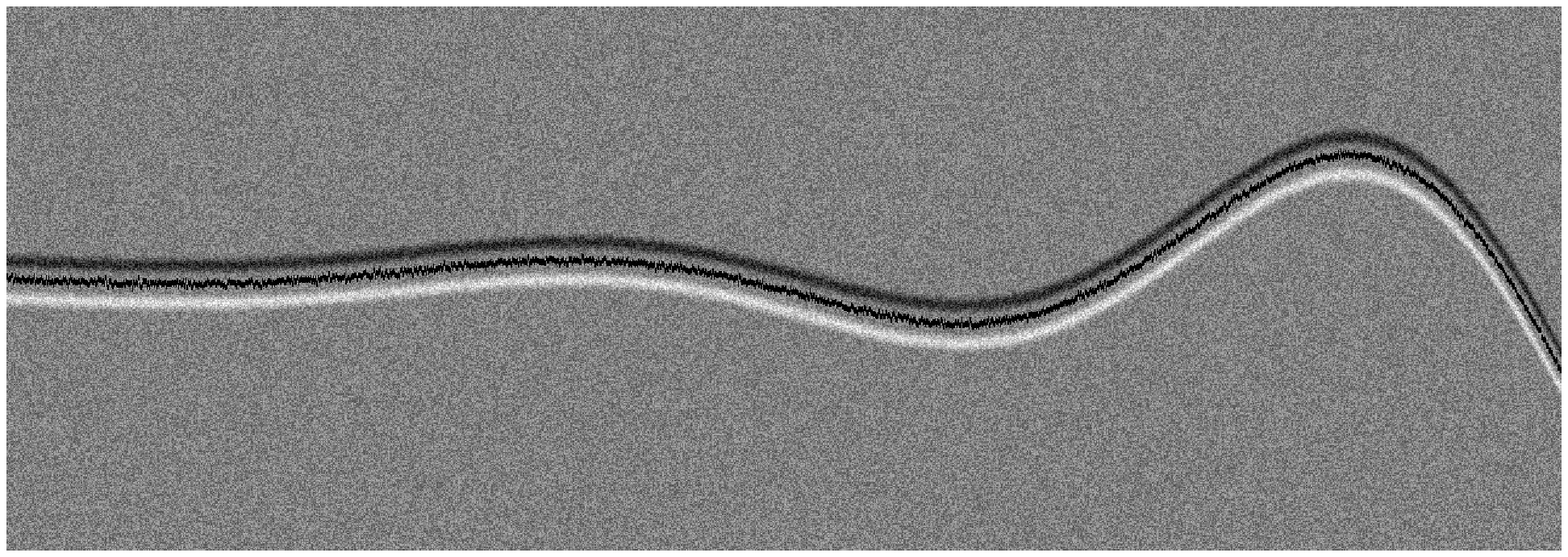}
\end{tabular}
\\
\begin{tabular}{c}
{\scriptsize   (\thefigure g) -0dB noise added (morphological analysis)} \\
\includegraphics[width=.95\columnwidth]{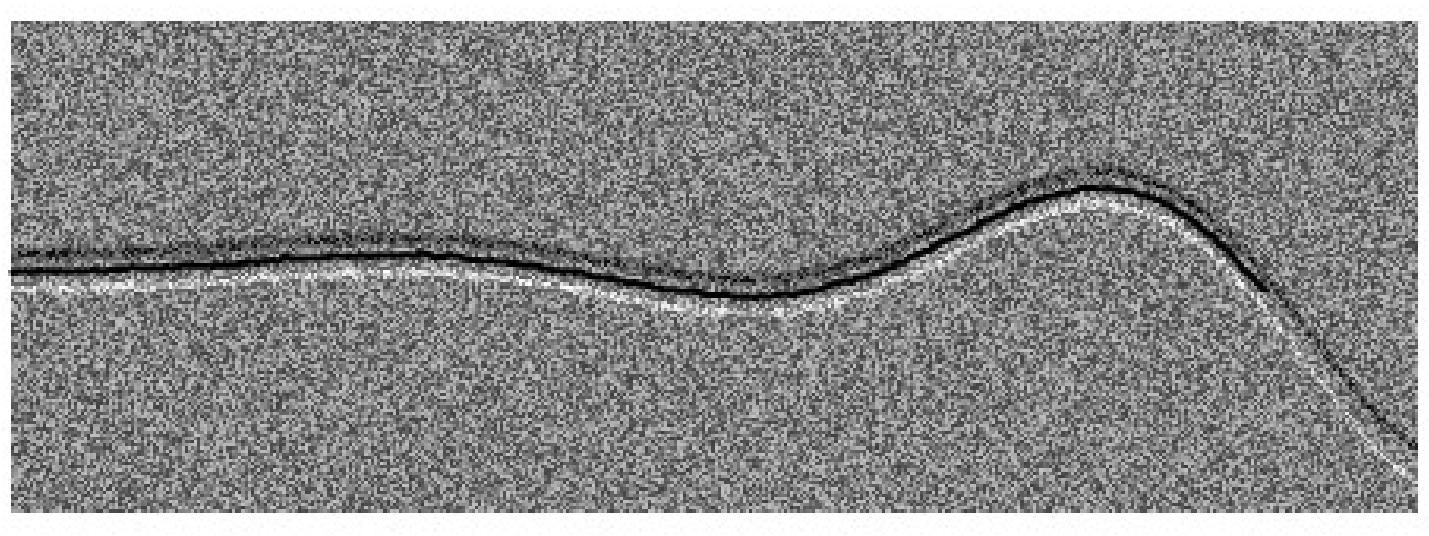}
\end{tabular}
&\hspace*{-1.cm}
\begin{tabular}{c}
{ \scriptsize (\thefigure h)   -0dB noise added (crosscorelation analysis)} \\
\includegraphics[width=.95\columnwidth]{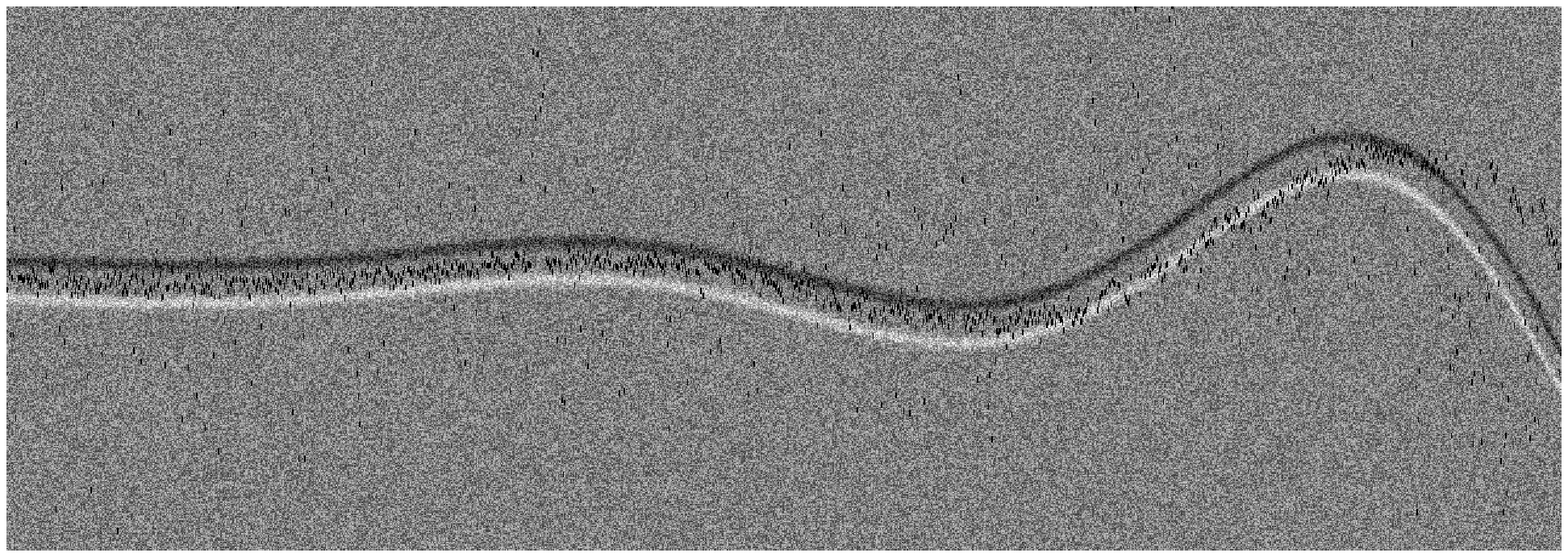}
\end{tabular}
\\
\begin{tabular}{c}
{ \scriptsize  (\thefigure i) +3dB noise added (morphological analysis)} \\
\includegraphics[width=.95\columnwidth]{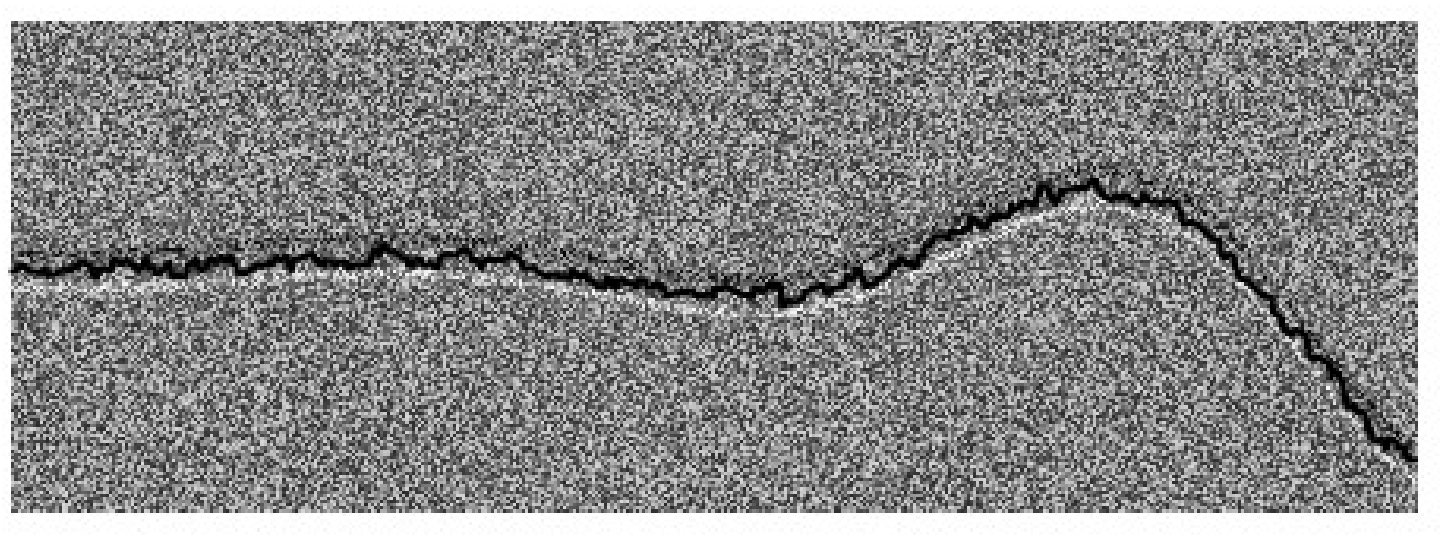}
\end{tabular}
&\hspace*{-1.cm}
\begin{tabular}{c}
{\scriptsize   (\thefigure j)  +3dB noise added (crosscorelation analysis)} \\
{\bf The method fails}
\end{tabular}
\end{tabular}
\addtocounter{figure}{-1}
\caption{Influence of signal to (white) noise ratio on the estimated   median line: comparison
  between morphological (left column) and crosscorrelation analysis (right).\label{f:synth_jet_noise_influence}}
\vspace*{-0.25cm}
\end{figure*}

Since a real image is always noisy, it is worth testing the influence
of noise on the image analysis results. To do this, white noise has
progressively been added to the original synthesized image. This
gives some insight on how algorithms behave when background noise is
increasing and contrast is decreasing. Results are presented in figure \ref{f:synth_jet_noise_influence} for different signal to noise ratio.


The morphological approach (left column) appears to estimate successfully 
the median line on the whole image length until $-3dB$ of noise. When
the noise ratio is increased to $0dB$, the median line is still estimated
but not until the right border of the image. In the noisiest case
($+3dB$) the range on which the median line would have been estimated
would be even shorter. To show how to prevent this, we have slightly
modified the parameters of the analysis: the empirical constant multiplying the Otsu binarisation threshold is increased up to 1.5 for the upper half jet and decreased down to 0.5 for the lower half jet. Moreover, the radius of the structuring element for the closure
has been divided by two. This allows to track the median line on the
whole length by manual adjustment of the analysis parameters, but the estimation is somehow more crude.

The crosscorrelation method gives reliable results for moderate levels of noise but shows more sensitivity than the morphological method for the noisiest images. However, real Schlieren images obtained in our experiment are less noisy than the extreme case tested here.

\section{Results comparison \label{s:results}}

\subsection{Data analysis}

\begin{figure}
\begin{center}
\includegraphics[width=.9\columnwidth]{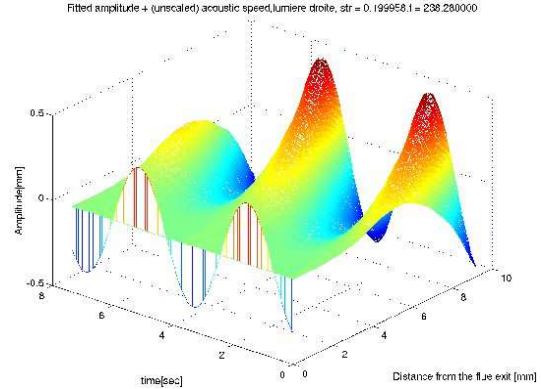}
\end{center}
\caption{Fitting of a sinusoidal model
  (cf. Eq. (\ref{e:3D_sinus_fit})) on the data plotted in figure \ref{fig:Position_mesh_morpho}. The wave superimposed at the flue exit corresponds to the perturbating acoustic velocity. It is plotted for phase reference.}
\label{fig:Position_fitted_mesh_morpho} 
\end{figure}

Every column of the images oscillates in time at the same frequency of the excitation. It is therefore possible to fit a sinusoid of that frequency to each column, and obtain the amplitude $\vert Y(f)\vert$ and phase $\angle Y(f)$ of the fitted jet transverse displacement:

\begin{equation}
Y(f) = \frac{1}{N}\sum_{k=0}^{N-1}X_k e^{-j k \frac{2\pi f}{f_s} },
\label{e:3D_sinus_fit}\end{equation} 
where $f_s$ is the sampling frequency, $N$ is the number of images in
the sequence, $X_k$ is the jet deflection for that column at image
number $k$.
These values $Y(f)$ are used to recreate a fitted version of the jet position as
shown in figure \ref{fig:Position_fitted_mesh_morpho}.

In a linear description of the jet (\cite{rayleigh45}, \cite{mattingly71}), the perturbation is amplified while being convected downstream. The perturbation travels at a velocity about one half of the jet speed.
Inspired from this theory, the experimental data can be analyzed assuming a jet transverse displacement $\eta$ following (\cite{delacuadra2005}):
\begin{equation}
	\eta(x,t) = \Re \left( \eta_0 e^{\gamma x}e^{i\omega( t - x/c_p)} \right)
\label{e:expo_ampl}
\end{equation}
where $\eta_0$ represents a complex initial amplitude of the
oscillation, $\gamma$ is the spatial growing rate, $c_p$ is the
convection velocity and $x$ is the distance from the flue exit.

The growing rate $\gamma$ can be estimated by fitting an exponential
to the detected amplitude curve. The fitting is performed on a range
of distances from the flue exit which is chosen as follows: the
starting point corresponds to the minimal resolution obtained with the
two methods (that is half a pixel for the morphological approach)
whereas the last point considered is such that the oscillation
amplitude reaches $80\%$ of the maximum oscillation amplitude estimated
by the morphological method. This upper limit was found empirically to
correspond to the limit above which an exponentially growing model is
no more relevant.

The convection velocity $c_p$ of the
perturbation can be estimated as the slope of the linear fitting to
the phase (on the same fitting range as for the estimation of the
growing rate $\gamma$) as shown in figure \ref{fig:Fitted_exp} for the cross correlation method and the morphological method, in the case of a laminar jet (Re=500, left column), and for a turbulent jet (Re=3000, right column). 

\subsection{Methods comparison}

\subsubsection{Laminar jet}

\begin{figure*}
\center
\begin{tabular}{cc}
\includegraphics[width=1.01\columnwidth]{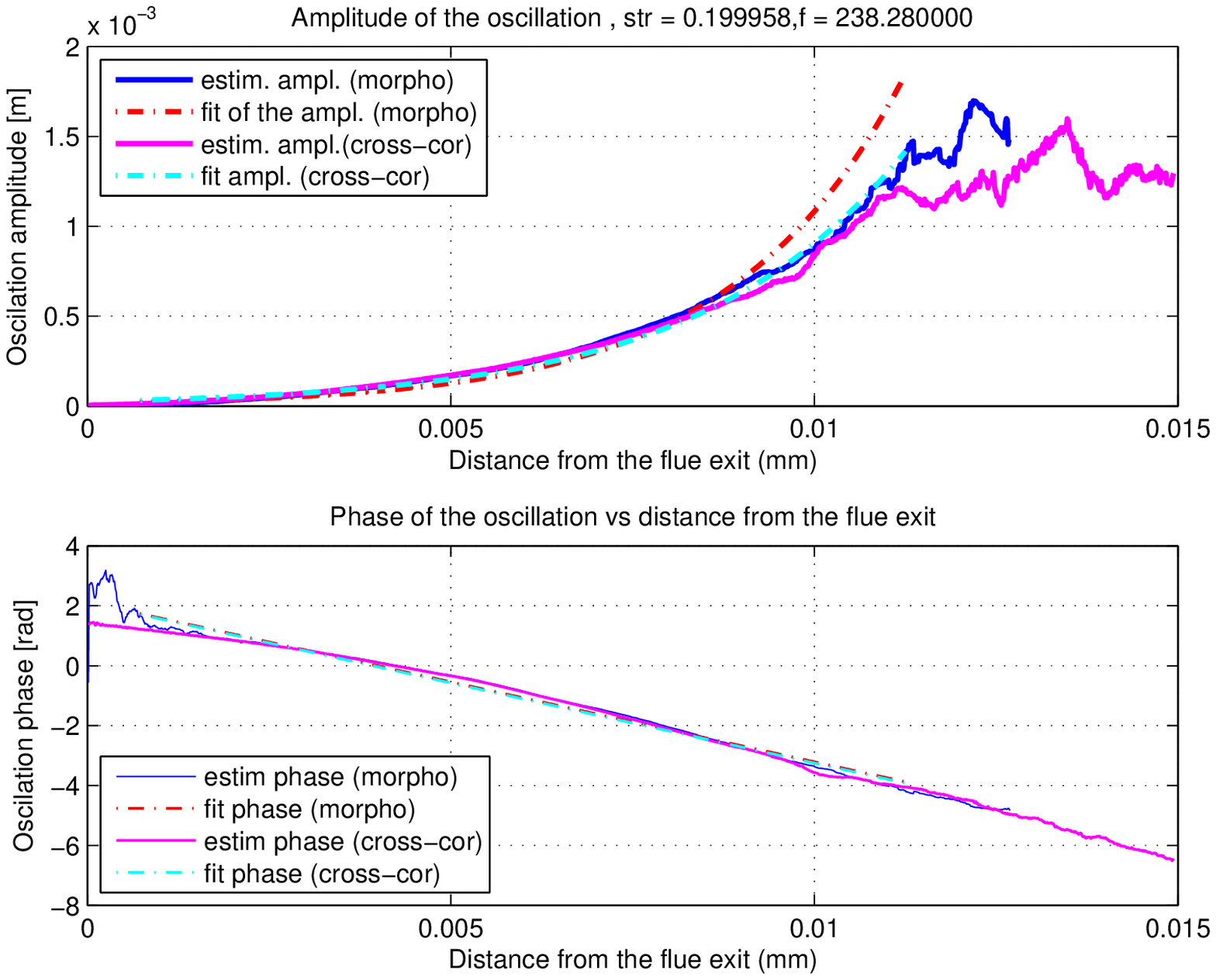}
&
\includegraphics[width=1.01\columnwidth]{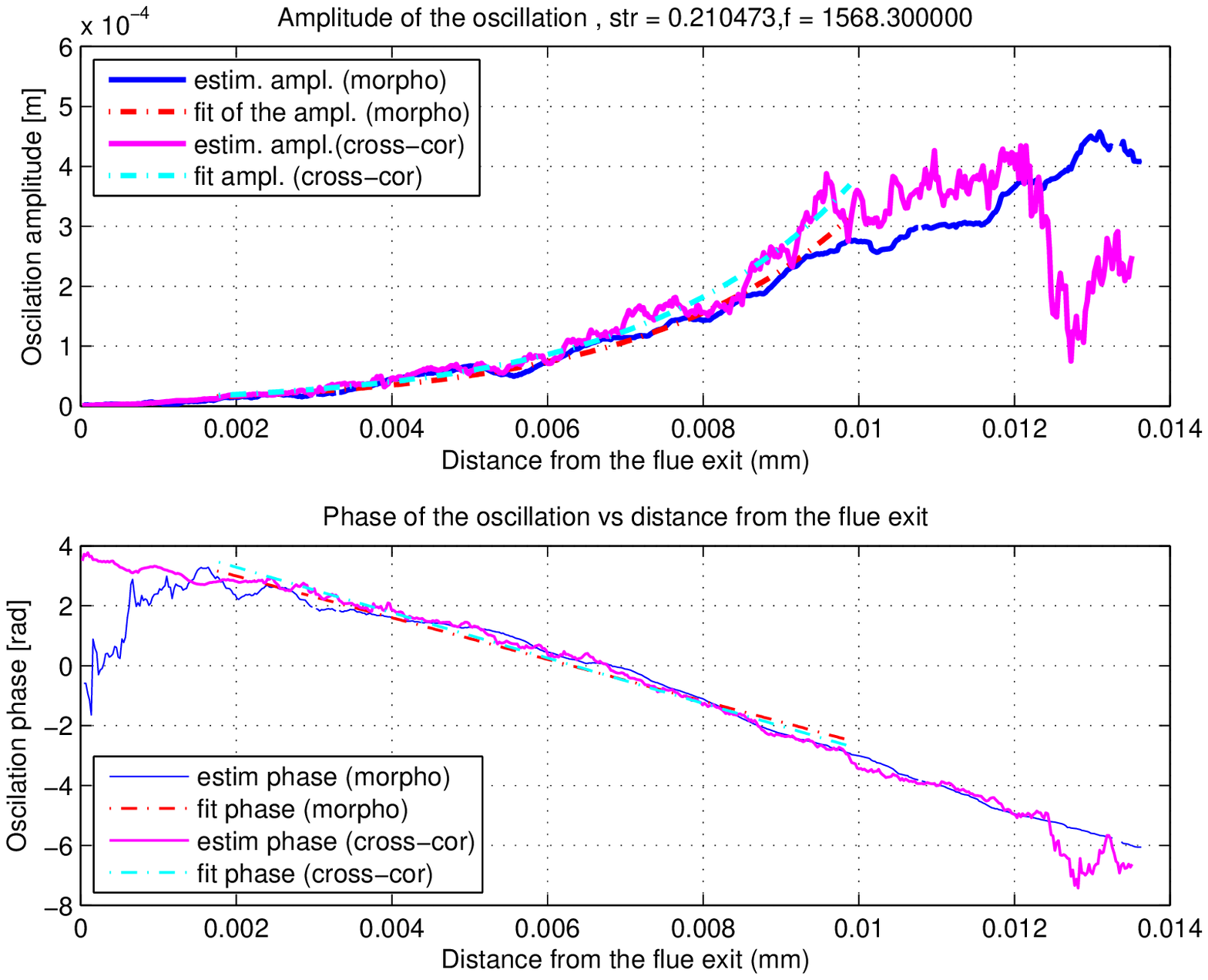}
\end{tabular}
\caption{Amplitude (top) and phase (bottom) of the jet transverse displacement using the cross correlation method (pink line) and the
  morphological method (green line). In the left column, the jet is laminar (Re=500, excitation frequency 238 Hz, i.e. St=0.195). In the right column, the jet is turbulent (Re=3000, excitation frequency is 1568 Hz i.e. St=0.21).} 
\label{fig:Fitted_exp} 
\end{figure*}

A laminar jet (Re=500) is first considered. We display in the left column of figure \ref{fig:Fitted_exp} amplitude and phase of the sinusoidal fit of the jet displacement. The two methods show very similar results: 
\begin{itemize}
	\item the oscillation amplitude follows roughly an exponential up to a distance of $x/h\approx11$ where $h$ is the thickness of the slit. 
	\item the phase shift of the oscillation decreases roughly linearly downstream from flue exit, corresponding to the delay induced by the convection of the perturbation on the jet at a somehow constant velocity. 
\end{itemize}


\subsubsection{Turbulent jet}
The case of a turbulent jet (Re=3000) is now investigated, for a
similar Strouhal number ($St=0.21$). Here again, the two methods give
similar results (see figure \ref{fig:Fitted_exp}, right column).  It should be
pointed out that, due to the lower relative acoustic excitation
amplitude in the turbulent case ($v_{ac}/U_j= 2.7 \, 10^{-3}$, compared to
$v_{ac}/U_j= 5 \, 10^{-3}$ in the laminar case), the transverse jet
displacement is much smaller on the right column of figure \ref{fig:Fitted_exp} than
on the left column ($\eta_{max}/h\approx 0.3$ compared to
$\eta_{max}/h\approx 0.8$). Analyses
in the right column of figure \ref{fig:Fitted_exp} therefore correspond to a very
severe case of a turbulent jet with very small transverse
displacement. Regarding the phase shifts, a downward bending is observed, as excepted
in the case of a rapidly slowing turbulent jet.

\subsubsection{Singular flow behavior}

For some conditions, the flow instability moves away from the simple exponential model described in the previous section. For laminar condition, this is the case for the roll-up of vortices  as well as for the development of varicose modes on the jet.

\paragraph{Roll-up of the jet:}

\begin{figure*}
\hspace*{-1.0cm}
\begin{tabular}{cc}
\begin{tabular}{c}
\begin{minipage}{\columnwidth}
\vspace*{-0.7cm}
\hspace*{1.0cm}\includegraphics[width=.9\columnwidth]{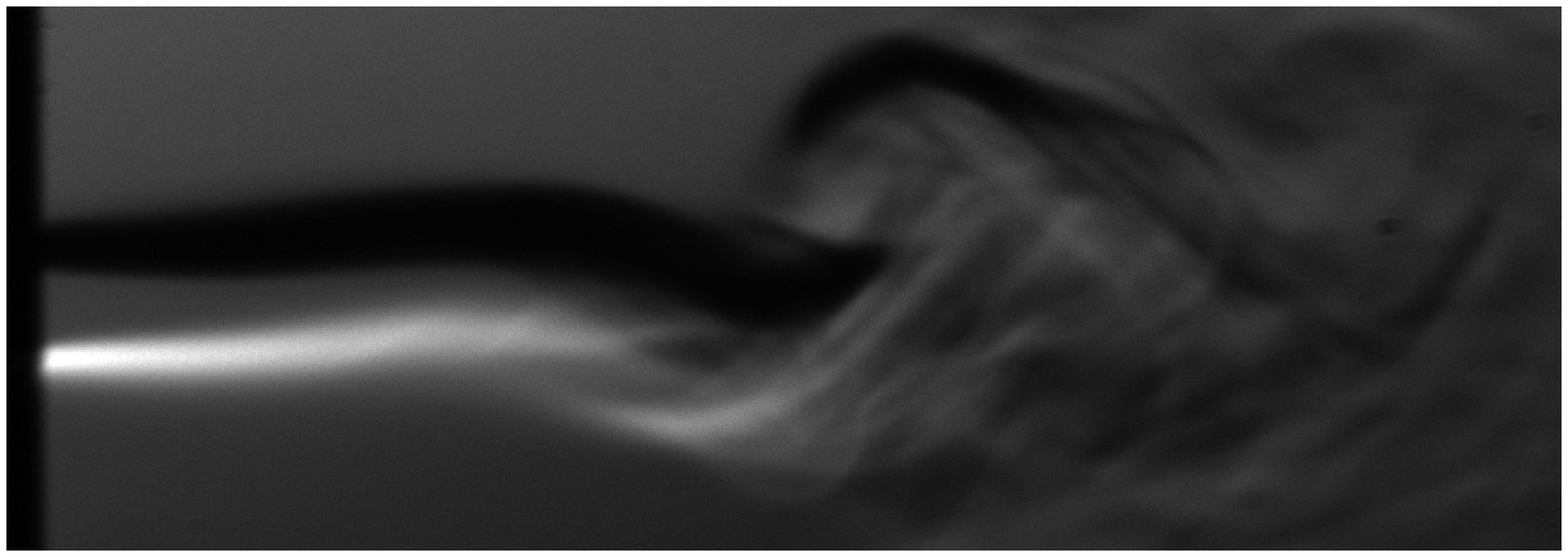} 
\end{minipage}
\\
\begin{minipage}{\columnwidth}
\includegraphics[width=1.11\columnwidth,height=8cm]{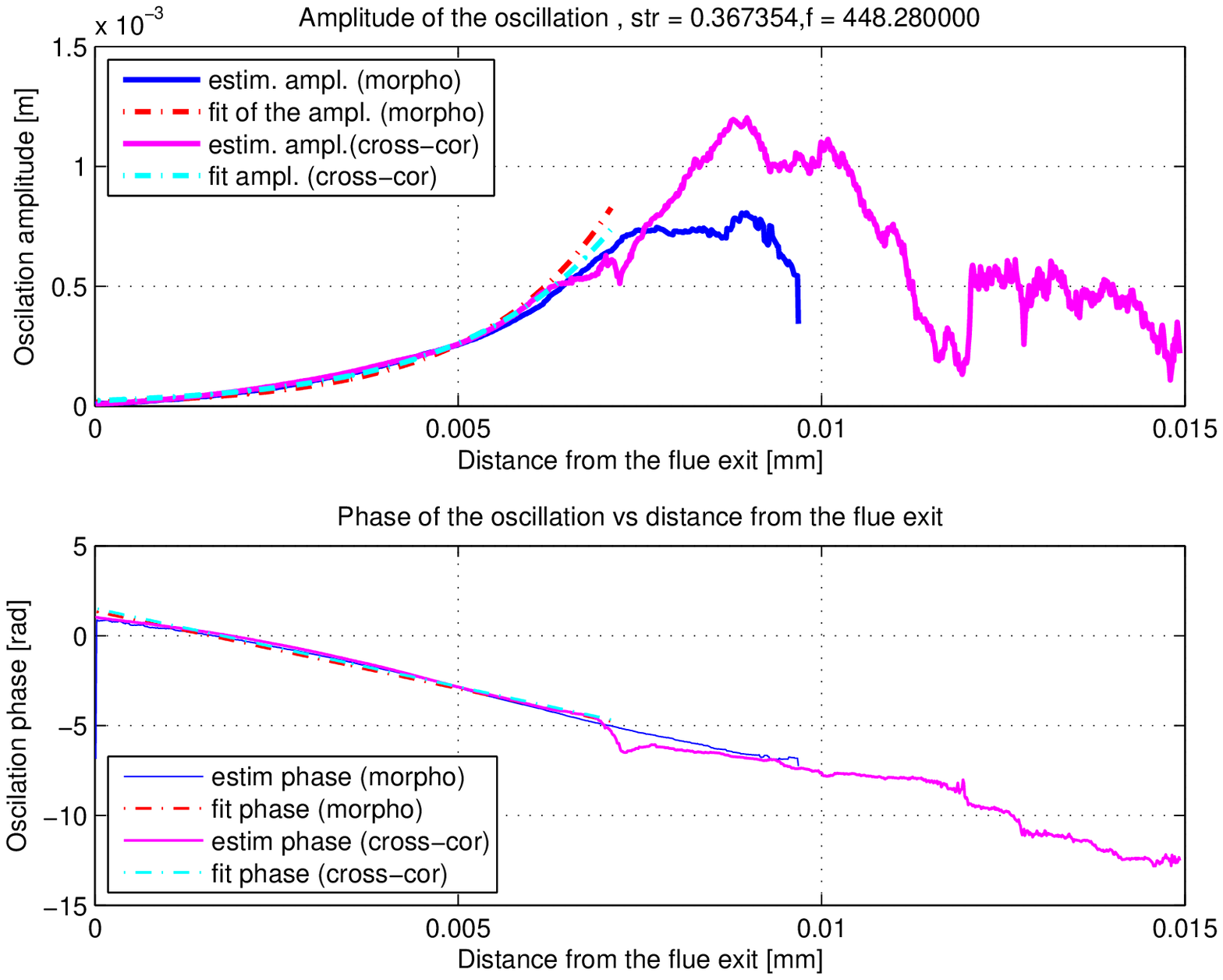}
\end{minipage}
\end{tabular}
&
\begin{tabular}{c}
\begin{minipage}{\columnwidth}
\vspace*{-1.05cm}
\hspace*{0.5cm}
\includegraphics[width=.9\columnwidth,height=2.5cm]{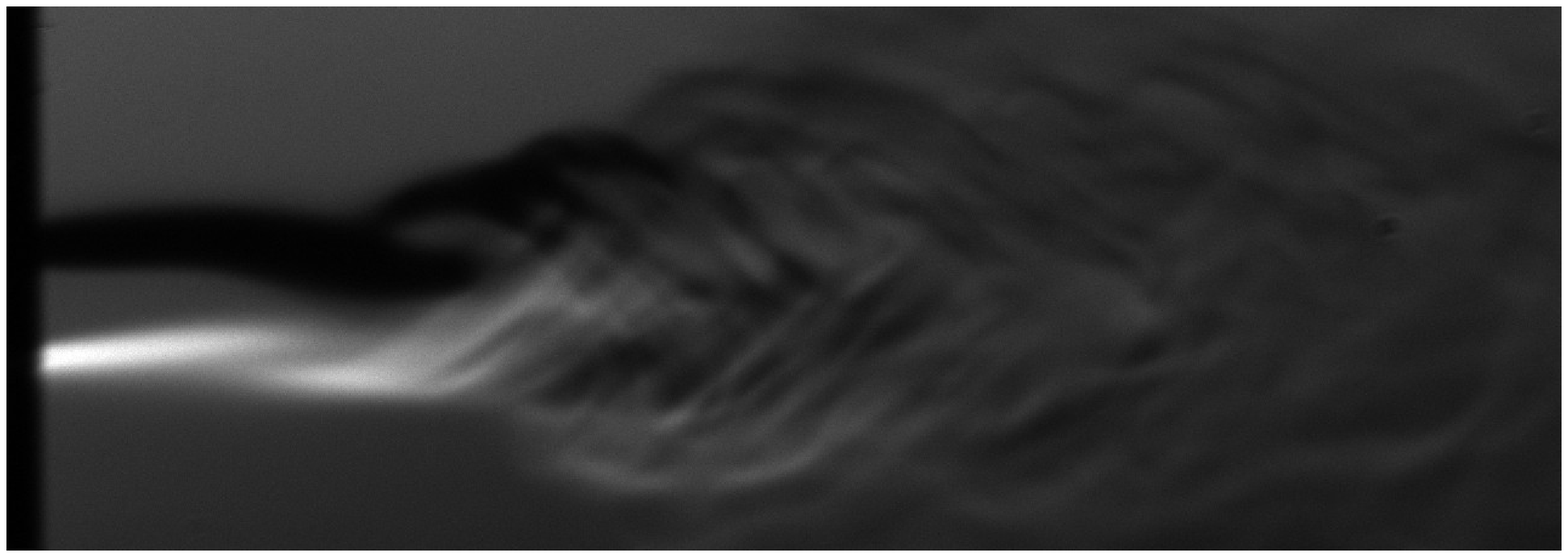} 
\end{minipage}\\
\begin{minipage}{\columnwidth}
\vspace*{.2cm}
\includegraphics[width=1.01\columnwidth,height=7.5cm]{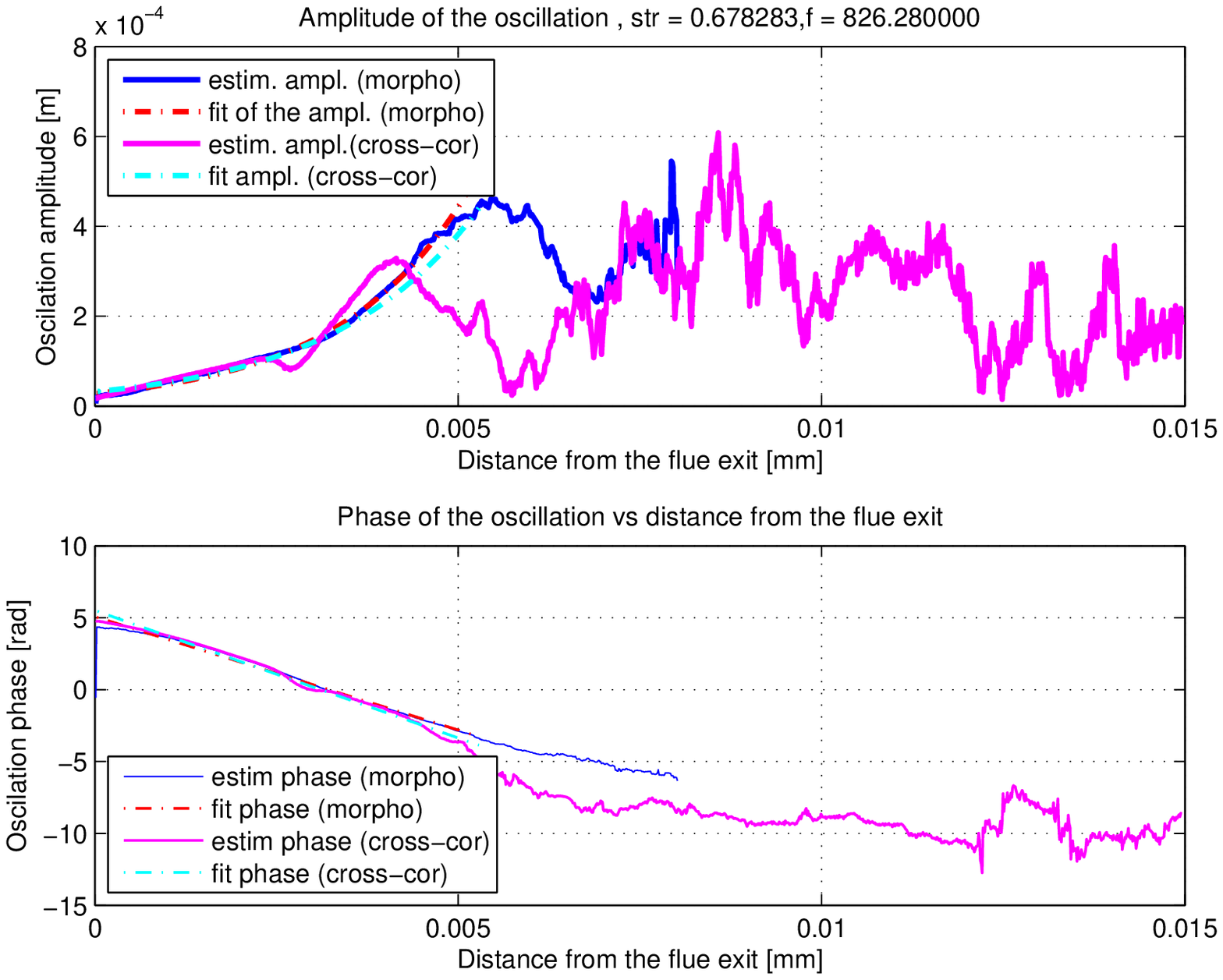}
\end{minipage}
\end{tabular}
\end{tabular}
\vspace*{-0.5cm}
\caption{Two illustrations of singular flow behaviours (one image of the sequence (top) and analysis results). Left column: roll-up of the jet (Re=500, St=0.36). Right column: development of varicose modes on the jet (Re=500, St=0.68).} 
\label{fig:singular_flow} 
\vspace*{-0.5cm}
\end{figure*}

In the case of strong jet amplification, the accumulation of the vorticity in the jet shear layers at the inflexion points induces the formation of vortices: as a result, the jet breaks into an alternated vortex street (\cite{holger77}). Depending on the Reynolds number, this vortex street may rapidly be decomposed by turbulence. Even if the cross correlation appears to track the jet transverse displacement further downstream than the morphological method, figure \ref{fig:singular_flow} (left column) shows  that both analysis methods used are affected by this vortex formation.

\paragraph{Developpement of varicose modes:}
Depending on the symetrical properties of perturbations on both shear layers of the jet, the jet oscillation can be sinuous or varicose. Sinuous oscillations are characterised by a transverse displacement of the jet while varicose oscillations correspond to an oscillation of the jet thickness (\cite{mattingly71}). Even if sinuous modes appear to be dominant  (\cite{verge94b}) in flute-like instruments, varicose contributions can sometime be observed related to fluctuations in the jet velocity. In the case of the presence of varicose oscillation on the jet (see figure \ref{fig:singular_flow}, right column), the morphological method appears to track the jet median line further downstream than the cross-correlation method.

A partial conclusion of these examples above is that, in both cases, the bias between results of the two methods is an indication for singular flow behavior.

\section{Selected results on the jet behavior  \label{s:disc}}
Two applications of the proposed methods are presented within this section. 

\subsection{Influence of the Reynolds number on the convection velocity}

\begin{figure}[h]
\center
\hspace*{-0.5cm}\includegraphics[width=1.01\columnwidth]{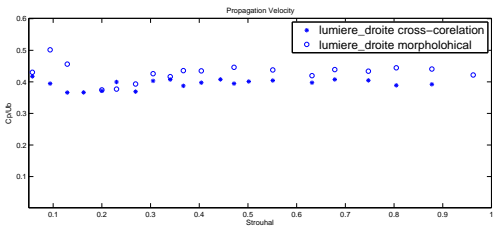}
\caption{Convection velocity of the perturbations on the jet as estimated through cross-correlation and morphological methods, laminar jet (Re=500).} 
\label{fig:summ_lam_cp} 
\end{figure}

\begin{figure}[!h]
\center
\hspace*{-0.5cm}\includegraphics[width=1.01\columnwidth]{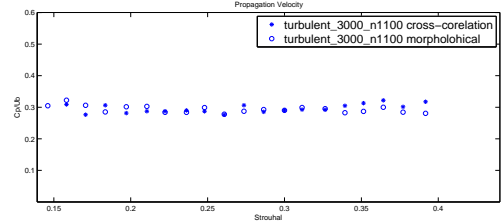}
\caption{Convection velocity of the perturbations on the jet as estimated through cross-correlation and morphological methods, turbulent jet (Re=3000).} 
\label{fig:summ_tur_cp} 
\end{figure}

The results obtained with cross-correlation and morphological methods
can be compared in a more general way by considering the evolution of
the estimated parameters as function of the frequency in the case of a
laminar (figure \ref{fig:summ_lam_cp}) and  a
turbulent (figure~\ref{fig:summ_tur_cp}) jet.
The mean convection velocity of the perturbation on the jet (defined as the mean slope of the phase shift) appears to be nearly independent of the frequency of oscillation. This is true for a laminar jet as well as for a turbulent jet. In the former case, $c_p/U_b \simeq 0.4$ and in the latter $c_p/U_b \simeq 0.3$. These results have to be compared with the Rayleigh ideal linear jet for which $c_p/U_b $ varies between $0.3$ and $0.5$ depending on the jet velocity profile. The values found are also in agreement with values suggested by other authors (\cite{mattingly71}, \cite{nolle98}).

A closer look at the evolution of the phase with the distance $x$ from flue exit indicates a higher convection velocity at short distances from flue exit (a downward bending is observed). This can be interpreted both in terms of the slowing down of the jet velocity and of the smoothing of the jet velocity profile, both induced by the viscosity. A fit of the phase with a power law reveals that the speed of the jet decreases as a function of $x^{-0.33}$ in the laminar case (mean value obtained from all the excitation frequencies available) and as a function of $x^{-0.53}$ in the turbulent case (mean value on all excitations frequencies also). From \cite{tritton88} (p133 and p323) the jet centerline velocity in the laminar case is expected to decrease as $x^{-1/3}$, while the decrease is much faster in the case of a turbulent jet since the jet centerline velocity follows $x^{-1/2}$. The fitted values appear to be very close to the theoretical values, even if it should be noted that the dispersion around the mean value is important, due to experimental problems for some frequencies. However, this approach appears promising and could be extended to tackle the detection of the transition from laminar to turbulence in a jet. 


\subsection{Influence of the channel geometry on the convection velocity}
Another application of the methods presented can be illustrated by the comparison of different flue channel geometries, as studied by Segoufin \cite{segoufin2004a}. Convection velocities of perturbations on jets issuing from long (27mm) and short (1mm) channels are compared. Both channels show the same thickness $h=1$mm. In the case of the short channel with squared exit, the convection velocity is clearly higher than in the other cases, confirming the observations done by S\'egoufin \cite{segoufin2004a} in the context of edge-tones. This can be interpreted, as suggested by S\'egoufin, in terms of the boundary layers' thickness of the jet, which becomes thinner in the case of short channel. The present experiment was carried using the channels also used by Segoufin, and our results strongly correlate with Segoufin's results and interpretation.
\begin{figure}[!h]
\hspace*{-0.25cm}\includegraphics[width=1.1\columnwidth]{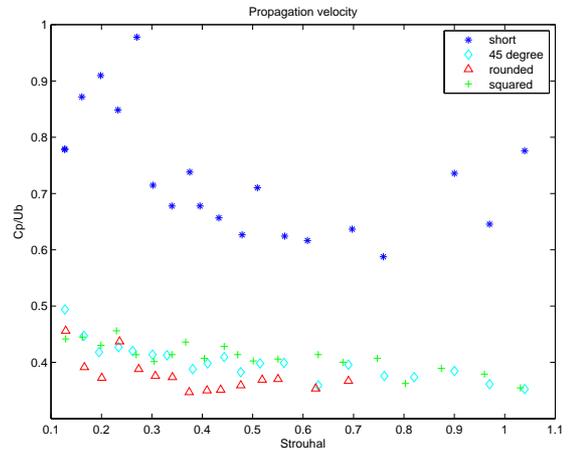}
\caption{Dimensionless velocity, flue exit short (dashed), squared (dashdot), rounded (solid) and 45 degrees (dotted).}  
\label{fig:vel_toutes}
\end{figure}

\section{Conclusion}

Compared to hot-wire anemometry, flow visualization allows a non intrusive measurement as well as a clear identification of the jet as composed of fluid particles issuing from the flue channed (streaklines,  \cite{hama62}). On the other hand, flow visualizations does not give direct information on the velocity field (streamlines) and requires the use of a gas of different light refraction than that of the surrounding air.

The images obtained by the Schlieren method are extremely sensible to variations on the setup. The position of the blade, the aperture of the diaphragm, the alignment of the optics, the saturation curve of the acquisition software, the vertical and horizontal position of the jet, etc. are some of the parameters that needs to be adjusted for each experiment. Small differences on these parameters produce image variations. Therefore, even under a attentive care, images obtained from any two experiments are different. 

Both methods allow to estimate fairly well the amplitude as well as
phase evolution of the perturbation. The main difference is that
the cross-correlation method performs an inter-image analysis while
the morphological method is an intra-image technique. Moreover, to achieve its best results the morphological method needs to be manually calibrated for each sequence of images while the cross-correlation method can be completely automate. The resolution of the detection is restricted to one pixel in the morphological method while continuous for the cross-correlation. Turbulent jet and noisy images seem to be better handled by the morphological method while the cross-correlation is preferred for laminar jets.

The slowing down of the jet can be quite accurately evaluated, and appears to follow the theoretical values expected for both laminar and turbulent jets.

The two methods we developped give very close results in terms of jet amplification and convection velocities for a wide range of control parameters. Together with the interpretation of results presented by Segoufin using the same channels, this indicates that both methods can be reliable and useful to the study of jet oscillation in edge-tones and flute-like instruments.

\paragraph{Acknowledgments :} The study presented in this paper was lead with the support of the French National Research Agency \textsc{anr} within the
\textsc{Consonnes} project.



\end{document}